\documentclass[journal]{IEEEtran}
\usepackage{amsfonts}
\usepackage{ifpdf}
\usepackage[justification=centering]{caption}
\usepackage{cite}
\ifCLASSINFOpdf
\usepackage[pdftex]{graphicx}
\else \fi
\usepackage[cmex10]{amsmath}
\usepackage{array}
\usepackage{mdwmath}
\usepackage{mdwtab}
\usepackage{eqparbox}
\usepackage[tight,footnotesize]{subfigure}
\usepackage{algorithmic}
\usepackage{algorithm}
\usepackage{amsmath}
\usepackage{epsfig}
\usepackage{ae}
\usepackage{amssymb}
\usepackage{times}
\usepackage{amsmath}   
\usepackage{booktabs}
\usepackage{color}
\usepackage{bm}
\usepackage{multirow}
\usepackage{epstopdf}
\usepackage{stfloats}
\usepackage{subfigure}
\usepackage{float}
\usepackage{url}

\begin{document}

\title{Full-Life Cycle Intent-Driven Network Verification: Challenges and Approaches}
\author{Yanbo Song$^\dag$, Chungang Yang$^\dag$, Jiaming Zhang$^\dag$, Xinru Mi$^\dag$, Dusit Niyato$^\ddag$
\thanks{$^\dag$Y. Song, C. Yang, J. Zhang and X. Mi are with the State Key Laboratory on Integrated Services Networks, Xidian University, Xi'an, 710071 China. (emails: songyanbo94@163.com;  guideyang2050@163.com; jiaming0703@163.com; xinrum@163.com).}
\thanks{$^\ddag$D. Niyato is with School of Computer Science and Engineering, Nanyang Technological University, 639798 Singapore. (emails:  dniyato@ntu.edu.sg).}
}
\maketitle

\begin{abstract}
With the human friendly declarative intent policy expression, intent-driven network can make network management and configuration autonomous without human intervention. However, the availability and dependability of these refined policies from the expressed intents should be well ensured by full-life cycle verification. Moreover, intent-driven network verification is still in its initial stage, and there is a lack of full-life cycle end-to-end verification framework. As a result, in this article, we present and review existing verification techniques, and classify them according to objective, purpose, and feedback. Furthermore, we describe intent verification as a technology that provides assurance during the intent form conversion process and propose a novel full-life cycle verification framework that expands on the concept of traditional network verification. Finally, we verify the feasibility and validity of the presented verification framework in the case of an access control policy for different network functions with multi conflict intents.
\end{abstract}
	
	
\section{Introduction} \label{section:1}	
Software-defined networking (SDN) is distinguished by programmability, flexibility, and decoupling of control and data planes. However, due to the growing network scale and business diversity, network management becomes more complex and challenging. Millions of forwarding rules dictate how vital devices behave in the network. Incorrect policies or configurations can result in network vulnerabilities that lead network outages, routing oscillations, and forwarding black holes, further impairing network availability and dependability \cite{zhang2020network}. Most network operators use low-level interfaces to program the network, which is practically inefficient and error-prone. It has been proven that the functionality of programmable networks cannot be implemented unless a high-level abstraction policy is provided to the users \cite{8256016}. Intent-driven network (IDN) promises to fill this gap by providing a simple, yet expressive high-level abstraction policy over the network controller. This abstraction policy hides unnecessary details of the underlying infrastructure from users and allows them to customize network configuration using human readable intents.
	
IDN is a novel network paradigm that has gained significant interest from industry and academia. An intent is an abstraction declaration of what applications require from the network. It is composed of a set of primitive ``verb'', each describing a specific but high-level operation \cite{pang2020survey}. High-level abstraction policies decrease the need for specialized expertise. However, network devices cannot directly comprehend and enforce them. There are certain differences between the semantics contained in the intent and the parameters of the physical network. For example, an intent ``to establish a high-speed link from node A to node B'' is with unclear semantics. Because there exist ``distortion'' and mismatch between service provider-defined and user-defined high-speed network. IDN must thus be able to translate intents into more detailed lower-level rules, and this process is known as intent translation. Each translation adds some ``distortion" due to lower-level constraints and hence the intent must be checked at each step to ensure that it stays accurate throughout the continuous translation. Intent is diverse and random, which can lead to uncontrollable network issues like intent conflicts among applications. As a result, intent verification is required. However, there is a lack of evaluation of intent execution effect and the verification of the intent's feasibility and validity is separately implemented.

IDN eliminates the inefficiencies of conventional network management and decreases the risk of misconfiguration by automatically converting abstraction intents to detailed network configurations. IDN contains multiple key functions: \emph{Intent Representation, Intent Verification, Policy Mapping, and Situational Awareness}. On the one hand, verification does exist in traditional networks, it is a point-to-point verification and focuses on checking compliance with network policies and properties such as paths among network nodes. On the other hand, intent verification has to be structured with a full-life cycle that automatically achieves the goal of an intent and policy correctness. In particular, an intent can be characterized as natural language or graphic language, and there exist different kinds of verification from intents to policies \cite{jacobs2019deploying}.
	
The verification process incorporates formal methods, mathematical reasoning, computer languages, and networks. The verification approaches fall into two types. The first is conventional network verification techniques, which are often based on solvers and customized network tools. The other is a breakdown of conflicts based on policy consistency. However, there is no standardized definition of verification procedures and a full-life cycle view of verification. As a result, this article first provides a brief survey and classification of verification techniques. We present a framework for full-life cycle intent-driven network verification and develop a use case that employs the policy graph abstraction to resolve disputes and then configures policy in a simulation network. This article makes the following contributions:

	\begin{table*}[t]	
	\centering \caption{A brief survey of validity, feasibility, and joint verification techniques.} \label{TAB:tab2}	
	\renewcommand\arraystretch{1.5}
	\begin{tabular}{|m{1.6cm}<{\centering}|m{1.8 cm}<{\centering}|m{2.6cm}<{\centering}|m{2.1cm}<{\centering}|m{5.8cm}<{\centering}|m{1.1cm}<{\centering}|}
		\hline
		Type                                      & Solutions    & Theory and Methodology           & Verifi Object    & Characteristic                                                                & Network \\ \hline
		\multirow{3}{1.8cm}{\centering Feasibility Verification} & PGA, Janus \cite{abhashkumar2017supporting}   & Policy Graph Abstraction                & Network Function & Provide an intuitive graph abstraction to express and compose policies.     & IDN     \\ \cline{2-6}
		& LUMI \cite{jacobs2019deploying}        & Nile \cite{jacobs2018refining} and Merlin  \cite{li2020modular}                 & Natural Language & Continuous   feedback to improve the accuracy of information extraction but fixed Priority Policy.                       & IDN     \\ \cline{2-6}
		& Evian   \cite{mahtout2020using}     & Resource Description Framework    & Natural Language & Intentional presentation platform with natural language interaction.                                                     & IDN     \\ \hline
		\multirow{7}{1.8cm}{\centering Validity Verification}  & ATPG    \cite{zhao2017troubleshooting}        & Probes and Header Space Analysis & Path             & Point   out the dynamic verification, and generates fewer test packages.      & SDN     \\ \cline{2-6}
		& SERVE  \cite{zhao2017troubleshooting}      & Probes                           & Rules            & Verification of data plane, less use of probes.                                                                          & SDN  \\ \cline{2-6}
		& Pingmesh \cite{zhao2017troubleshooting}    & Ping                             & Path             & Single Packet .                                                                                                          & DCN     \\ \cline{2-6}
		
		& VeriDP  \cite{zhang2020network}     & Packet Tag                       & Path             & Forwarding behavior verification, but need to modify the switch hardware and software  .                                 & IP      \\ \cline{2-6}
		
		& VeriFlow  \cite{horn2019precise}    &     Mathematical modeling            EC               & Rules            & Real-time network verification.                                                                                          & SDN   	\\ \hline
		\multirow{2}{1.8cm}{\centering Joint Verification}
		& Monocle \cite{perevsini2018dynamic} & Agent        & Configuration    & Express the logic of the switch forwarding table as a boolean satisfiability problem. & IP      \\ \cline{2-6}
		& Mineseweeper \cite{beckett2017general} & BatFish and SMT Formulate        & Configuration    & Encode   all possible packet behavior within the network using first-order logic, but solve all constraints as a whole. & IP      \\ \cline{2-6}
		& Epinoia   \cite{wang2021epinoia}   & Graph Abstraction        & Network Function & Extends the intent specification in PGA and supports incremental checks.                                                        & IP      \\ \hline
	\end{tabular}
\end{table*}
	
\begin{itemize}
	\item Due to the lack of clear definition and classification of IDN verification, we define the IDN verification as a full-life cycle verification from a language translation standpoint, and present a brief survey and a classification of the current verification technology.
		
	\item As there is no unified design of the IDN verification framework, we therefore present an intent-driven network full-life cycle verification framework, which is with both intents and network status in the policy graph abstraction.
		
	\item To evaluate the full-life cycle verification framework, we implement a full-life cycle to black the virtual network function orchestration and the packet arrival rate verification.
\end{itemize}

The remainder of the paper is organized as follows. We first overview the concept and classification of intent verification in Section \ref{section:2} and Section \ref{section:3}. Then, we propose the full-life cycle intent-driven network verification framework and conduct simulations and evaluations to verify the effectiveness of the proposed framework in Section \ref{section:4}, followed by future research and concluding remarks in Section \ref{section:5} and \ref{section:6}.
	
\section{An Overview of Intent-Driven Network Verification} \label{section:2}

IDN aims to provide a more natural and intuitive network administration technique than conventional network management paradigms. An example of a general configuration and an example intent \textcolor{black}{is} presented as follows:
\[\begin{array}{l}
	Configuration:\\
	If\_(match(srcip=ZoneB,{\rm{ }}dstport=80,{\rm{ }}dstip=ZoneA))\\
	Intent:	\\
	The \ {\rm{ }}traffic \ {\rm{ }} from{\rm{ }} \ ZoneB{\rm{ }} \ to{\rm{ }} \ ZoneA\ is{\rm{ }} \ allowed{\rm{ }} .
\end{array}\]
	
The $Configuration$ contains specific information about what the switch must do (i.e., match the destination IP address and port, and then forward the packet). Meanwhile, the $Intent$ only describes a desire (i.e., traffic from Zoom B to Zoom A is allowed). Since the intent only describes the abstract desire and lacks many configuration information, it is necessary to complete the details and verify the correctness of the process.

We first classify, discuss, and compare the existing network verification techniques. In general, IDN expands the scope of verification, because only high-level abstraction policies have consistent verification problem. Therefore, feasibility verification often occurs in IDN. The verification techniques in conventional networks, such as IP and SDN, focus on the effectiveness of the feedback from underlying devices. Next, we elaborate some details of the related works. The characteristics of each schemes related to intent verification are shown in Table \ref{TAB:tab2}. We review the state-of-the-art in feasibility verification, validity verification and joint verification. Feasibility indicates whether the policy can run in the network, validity is an attribute that determines whether a policy meets certain requirements, and joint verification combines the above two attributes.

\subsection{Feasibility Verification}
	
	Feasibility verification ensures that the policies are executable in the network, conflict-free between policies (internal), and conflict-free between policies and underlying constraints (external).
	
\subsubsection{Verification with graph } The northbound interface (NBI) of the IDN allows users express their intentions, and avoids conflicts between intents \cite{pang2020survey}. The NBI enables intent conflict resolution before it is issued to the SDN controller. The intent conflict handling capability is more challenging when a vast number of policies are issued. For instance, the policy graph abstraction (PGA) provides a simple and intuitive graphical interface that is similar to how network managers typically visualize their policies on a whiteboard. Through the graph editor and graph composer, the output is a conflict-free policy graph. On the basis of PGA, Janus system expands the graph combination to dynamic policy, maximizing the number of configured policies and minimizing the number of path changes caused by either intrinsic dynamics in policies or due to policy churn \cite{abhashkumar2017supporting}.

\subsubsection{Verification with natural language} In addition to graphs, an intent can be expressed in natural language. Therefore, several works have focused on natural language processing (NLP). In the LUMI scheme, information extraction by named entity recognition allows collecting feedback from operators and incorporating it into the information extraction process, which is continuously learned or trained to improve the accuracy of information tagging \cite{jacobs2019deploying}. The LUMI scheme analyzes the conflicts arising from Nile after confirming a successful extraction of entities \cite{jacobs2018refining}. Nile is a scheme for learning network behavior expressed by the operator while providing a user-friendly interface to assist in the verification process of intent concretization. The translation process consists of three phases: entity extraction, intent translation, and intent deployment. For identifying the intent, the Evian client uses NLP and machine learning to build an intelligent bot that can have similar human conversations with users \cite{mahtout2020using}. The bot will use multiple conversations with users in English to gather all requirements about their network use cases.

	
\subsection{Validity Verification}
	
The advantage of validity verification is to ensure the policy is able to achieve the requirements for the network, including the efficient translation of policy to profile (offline) and the \textcolor{black}{effective} execution of policy to forwarding behavior (online).
	
\subsubsection{Verification using probe}
The southbound intent verification is achieved by capturing the configuration of the network data plane or by collecting real traffic as feedback. The control plane adjusts and modifies the policy formulation based on the feedback to achieve consistency before and after policy devolution. For complex network debugging problems, dynamic policy verification methods are gradually replacing static verification. ATPG is an automated and systematic approach to \textcolor{black}{network testing and debugging} as a transparent agent deployed in the middle of the control plane and data plane \cite{zhao2017troubleshooting}. At the same time, ATPG reads the switch configuration and generates device-independent models, sends test packets at regular intervals to detect faults, and designs fault location mechanisms. The type of verification is validity verification, and the verification taxonomy refers to $R=R'$ static verification and $R'=F$ dynamic verification. SERVE is an SDN rule verification framework that can automatically identify data plane network problems \cite{zhao2017troubleshooting}. By modeling the network device as a stateful multi-root tree of pipeline processing, the number of probes used is reduced.
	
In data center networks (DCN), TCP packets sent and received by edge servers can continuously detect network conditions and performance issues, such as end-to-end delay. Moreover, anomalies in key performance data in the network can directly react and determine whether there are network problems. In the Pingmesh scenario, the above approach also provides data support for the definition and tracking of service level agreements \cite{zhao2017troubleshooting}.
	
\subsubsection{Verification using models}
Network model is an effective method to \textcolor{black}{evaluate} network policies by modeling the network state in the data plane; such as firewalls, load balancing, and other network functions, and then considering whether the network violates network policies based on the constructed model.

VeriDP is a proxy deployed between the control plane and the data plane \cite{zhang2020network}. VeriDP abstracts all rule configurations on the control plane into a path table. It tags data packets and checks the tag information of the data packets to see if the forwarding is correct. The practical deployment verifies that the VeriDP server is on the control plane and the data collection pipeline is on the data plane. However, \textcolor{black}{the switches require} hardware and software modifications and \textcolor{black}{are} not easily applied directly to the existing network.

Existing tools require fine-grained time scales for checking profiles and data plane states. Static analysis of the network data plane is performed offline, leading to problems such as not detecting or blocking errors when they arise during network operation. The VeriFlow layer is designed between SDN controllers and forwarding devices to obtain a snapshot of the network as it evolves \cite{horn2019precise}. Furthermore, by dynamically checking the validity of network invariants as each rule is inserted, modified, or deleted. To ensure a real-time response, VeriFlow introduces incremental algorithms to search for possible errors. The key technologies are mathematical modeling, fast rule checking, and analysis. \textcolor{black}{VeriFlow is deployed with a proxy between the control plane and the data plane without feedback from real traffic on the data plane. This type of verification is called validity verification.}
	

\begin{figure*}[t]
	\centering
	\includegraphics[width=0.8\linewidth]{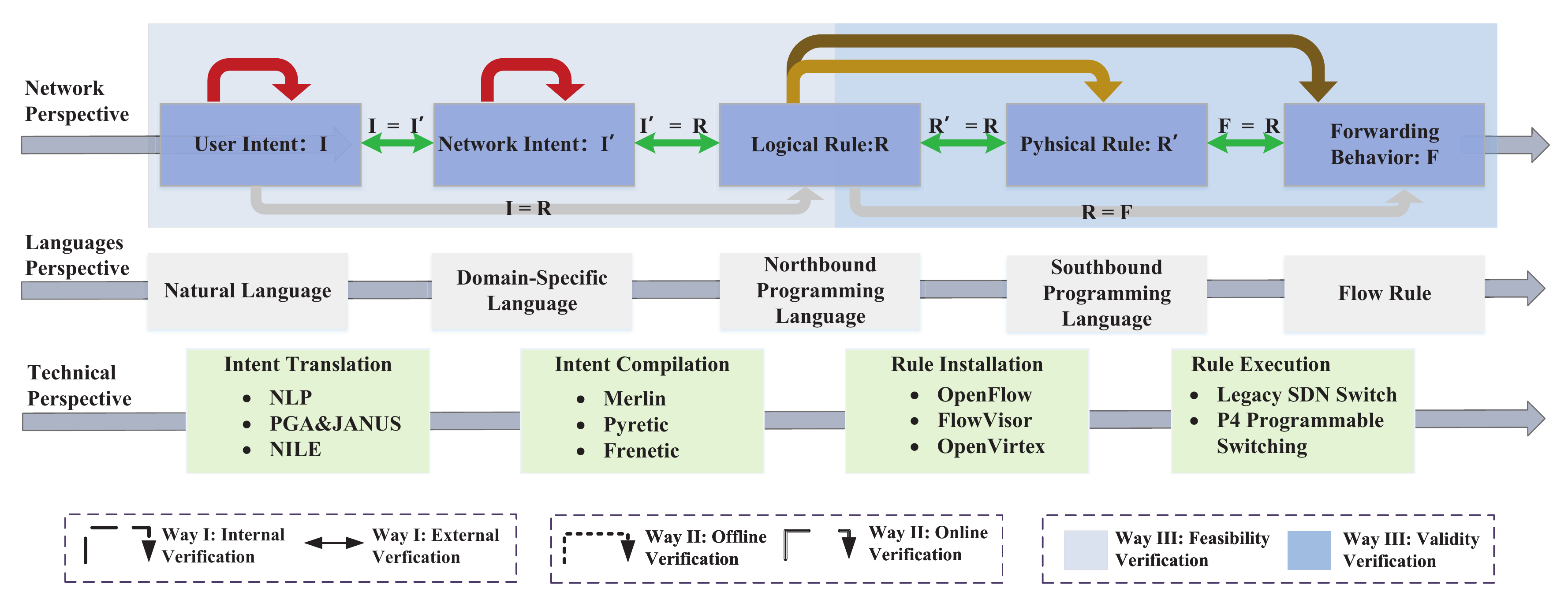}
	\caption{Different expression forms of intents are given from both the network and language perspectives, and the typical technical examples for translation are given between different intent forms. The colored arrows describe the three classifications of verification techniques.}
	\label{fig:intent-language}
\end{figure*}
	
\subsection{Joint Verification}

The purpose of feasibility verification and validity verification are complementary; feasibility verification ensures that the policies are executable, while validity verification ensures that the policy can satisfy the network's requirements; the combination of the two can produce better results.

Monocle addresses the inconsistency issue in policy as a result of complex network configuration and data plane \cite{perevsini2018dynamic}. The key technique of monocle is mathematical modeling, where the switch forwarding table logic is constructed by representing it as a boolean satisfiability problem, and probe packets check the practical switch behavior. Monocle is deployed as a transparent proxy between the control plane and data plane, again without real traffic feedback and in the form of the packets. Monocle is positioned as a layer between the OpenFlow controller and the switch. This design allows Monocle to intercept all rule modifications issued to the switch and maintain the expected flow table contents in each switch. After determining the expected state of a switch, Monocle can calculate the packet headers for rules to be enforced on that switch. Its validation types include feasibility and validity verification. To ensure that the network configuration and status derived from network automation match the administrator's specified intent, Epinoia has designed a network intent checker for stateful networks \cite{wang2021epinoia}. Epinoia expands the PGA-based on a unified network function model and gradually checks for intent violations within the network to reduce the impacts and costs of network changes. Minesweeper is a tool that verifies whether a network meets a variety of expected properties, including reachability or isolation between nodes, waypoints, black holes, bounded path length, load balancing, functional equivalence between two routers, and fault tolerance \cite{beckett2017general}. Minesweeper converts the network configuration file into a logical formula that captures the stable states to which network forwarding will converge, and these states are the result of interactions between routing protocols such as open shortest path first (OSPF), border gateway protocol (BGP), and static routing. Then minesweeper combines the constraints that describe the expected properties. If the combination formula is satisfiable, the network has a stable state. Otherwise, there is no steady state that violates the properties.
	
\textcolor{black}{The joint verification approach of feasibility and validity will be more comprehensive in terms of verification effectiveness than verification techniques that verify a certain property separately. It can be concluded} that IDN lacks a clear verification definition and classification. Therefore, the joint use of multiple verification techniques is required to guarantee the full-life cycle performance of network configurations. As a result, we define a full-life cycle verification definition to present full-life cycle of intent and the translation of intent language.

\section{Definition and Classification of Intent-Driven Network Verification}\label{section:3}
	
\textcolor{black}{
	Although the standard organization has a preliminary definition and verification classification, it is still insufficient to cover the entire IDN \cite{pang2020survey}. We define and classify verification based on the objective, purpose, and feedback, as well as characterize in different perspectives.
}
	
\subsection{Definition of Intent-Driven Network Verification}

	IDN is a revolutionary network paradigm in which intent is viewed as a collection of high-level abstraction policies. High-level abstraction policies reduce the need for specialized knowledge. However, they cannot be directly understood and enforced by network devices. Thus, IDN must be capable of translating intents into more precise lower-level policies, a process referred as intent translation. As illustrated in Fig. \ref{fig:intent-language}, the form of intent is continually changing and translating. Each translation introduces some ``distortion" due to lower-level limitations, therefore the intent should be verified at each stage to verify that it remains correct throughout the continuous translation. In IDN, the intent flow is expressed as follows:

\begin{itemize}
	\item \textbf{User Intent $I$ in Natural Language:} Natural language is the media via which users convey their intents about network functionalities and performance. The users who are unfamiliar with network can also communicate their intents in natural language.
		
	\item \textbf{Network Intent  $I'$ in Domain-Specific Language:} The domain-specific language is used to standardize the unstandardized natural language and hence improving the clarity of the network intents. Typically, the domain-specific language is constructed of tuples with specified names, such as \{domain, attribute, object, action, and result\}.
		
	\item \textbf{Logical Rule $R$ in Northbound Programming Language:} Northbound language refines intents by encapsulating them in practical network functions or algorithms. Because these network functions and algorithms act on the controller's logical view of the network and have not yet been constructed on the switch, they are referred to as logical rules.
		
	\item \textbf{Physical Rule  $R'$ in Southbound Programming Language:} Southbound languages like OpenFlow, sFLow, NetFlow, and simple network management protocol (SNMP) can be used for translating logical rules into physical rules that can be applied to network devices.
		
		
	\item \textbf{Forwarding Behavior  $F$ in Flow Rule}: Eventually, the intent will be translated into the network's forwarding behavior, which is determined by the flow table rules. The network device processes and forwards data packets based on the rules.
	\end{itemize}

The goal of IDN is to ensure that ``user intent" is translated into ``packet forwarding behavior." The user intent mentioned above can be user intent, network intent, logical rule, physical rule, and forwarding behavior. The ``rules" represent the ``intents" semantics. The intent translation process begins by adopting natural language and then converting high-level natural language into various levels of rules. As a result, the verification ensures that the semantics carried by the intent are preserved as much as possible during the translation process. The content to be verified at each stage differs. After the intent is converted into rules, the primary goal of the verification is to determine whether the policy implementation meets the expectations of the intent. Thus, full-life cycle verification ensures that the original semantics can be maintained between intents in any form, and the full-life cycle refers to the time between the generation of the intent and the end of the intent.	Therefore, \emph{Full-Life Cycle Intent Verification} can be defined as:
	
\textcolor{black}{\[F = R' = R = I' = I.\]}

Due to the limitation of verification technology, in practical networks, the current researches on verification don't distinguish all the conversions. Researchers tend to focus on one or several parts of the formula. NLP technology can be used to standardize natural language. PGA and Janus platforms resolve internal conflicts of intents through policy graph abstraction \cite{abhashkumar2017supporting}. The traditional SDN programming languages, such as Pyretic, Frenetic, and Merlin can be installed in the controller after being compiled \cite{li2020modular}. FlowVisor and OpenVirtex can check the correctness of the policy before and after the rule is issued, respectively \cite{han2016intent}. Researchers have achieved varying degrees of verification depending on their own technologies. They made a distinction between the control plane and the data plane. However, they lack the complete classify of verification process. As a result, we rearrange the classification of verification techniques according to the definition of verification.

\subsection{Classification of Intent-Driven Network Verification}
	
We summarize and classify the existing verification technologies according to the location in $F = R' = R = I' = I$. As shown in Fig. \ref{fig:intent-language}, we categorize verification technologies by where it occurs, whether there is feedback, and what purpose is.
	
\begin{itemize}
\item The verification techniques can be classified as ``Internal" and ``External" depending on the location of a verification object. Internal verification verifies multiple $Xs (X= I, R)$ in a layer, which is simpler to resolve at higher levels. External verification verifies the correctness of translation between layers, which relies more on underlying device feedback. The dedicated network model is insufficient to cover the full-life cycle intent verification. Since an intent may be expressed at several levels of abstraction, e.g., natural language, programming language, and graph, there may be overlap between the representations of $I'$ and $I$. The outcome of high-level abstraction translation needs to be checked only when the representation of $I'$ is at a very high degree of abstraction, like in a domain-specific language. Therefore, the top layer verification $R=I'$ should focus on internal consistency or conflict resolution to avoid a more severe impact on the network.
		
\item  The verification techniques can be further classified into ``Static" and ``Dynamic" verification, or ``Offline" and ``Online" verification, based on the feedback from the data \textcolor{black}{plan} during the verification process. Offline verification verifies the device profile of the data plane, focusing on $R=R'$. Online verification verifies the network status of the data plane by real-time collection, focusing on $R=F$ and whether the forwarding behavior in the network meets the policy requirements. The distinction is primarily in whether the data plane's real-time network state is collected.
		
\item According to the verification purpose, ``Feasibility" and ``Validity" are more general and comprehensive than the other two classifications. Feasibility verification ensures that policies are conflict-free with each other and the device constraints; Validity verification ensures the validity of configurations and forwarding behaviors. In addition, the conflict of high-level intents and policies needs to be considered; for example, the intents may come from different application requests. The objects targeted by the intent are shared resources, i.e., bandwidth, which will cause a conflict in the policy after intent translation. The conflict should be checked and dissipated before intents are finally executed.
		
\end{itemize}  	
	
Verification should focus on the full-life cycle of intent, including the entire verification process. A sound verification scheme should have a full-life cycle with feedback. Way III in Fig. \ref{fig:intent-language} is a more generic and inclusive classification scheme used to classify verification techniques.

\begin{figure*}[t]
	\centering
	\includegraphics[width=0.55\linewidth]{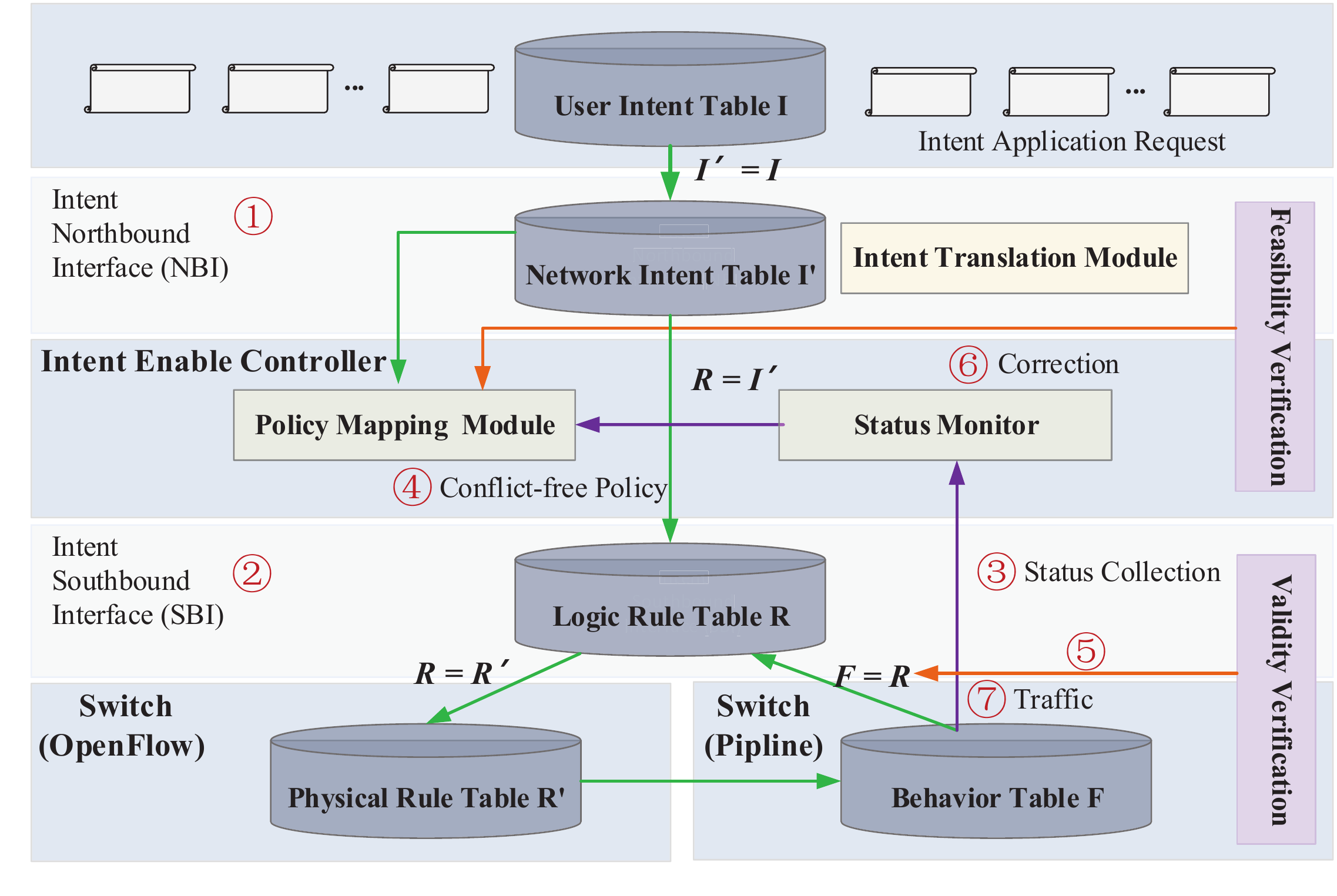}
	\caption{Full-life cycle intent-driven network verification framework. The verification module is deployed in the control layer, which is connected with the application layer and switches through northbound and southbound interfaces.}
	\label{fig:verfication-arc}
\end{figure*}
	
\section{A Full-Life Cycle Intent-Driven Network Verification Framework and Implementation} \label{section:4}

The intent verification is related to the stability and reliability of the whole network. Current verification techniques form independent modules for different aspects. As a result, we integrate these various aspects of verification techniques in this article. We propose a framework for intent verification in IDN, which contains the full-life cycle of $F = R' = R = I' = I$, and includes an application layer, an intent enable controller, an infrastructure layer, and two interface, as shown in Fig. \ref{fig:verfication-arc}.

The application layer requests a service, which could be presented as nature language $User\ Intent \ Table \ I $ or graph. User's intent expressed by the graph is already a standardized language $I'$, there is no need for $I = I'$ verification. However, possible conflicts among different intents still exist. The formalized intents by intent translation module are sent to the controller through NBI \textcircled{1}. Additionally, the intent verification module collects the intents through NBI and forms the $Network \ Intent \ Table \ I' $ composed of historical intents. When the controller receives an intent and implements a policy, which is generated according to both the intents and the network status. Status Monitor \textcircled{3} collects the network state. Policies should be combined with the feedback given by the intent verification module. If the policy fails to pass the verification module, the policy will be re-mapped \textcircled{6}. The conflict-free policies \textcircled{4} are sent down to the switches through the southbound interface (SBI) \textcircled{2} and forms $Logic \ Rule \ Table \ R $. After the policy is issued, the practical flow behavior of the network traffic \textcircled{5} is collected and reported to the intent verification module in real-time and form the $Behavior \ Table \ F$ \textcircled{7}. Then, according to the full-life cycle verification definition:

\begin{itemize}
	\item The feasibility verification of $I$ and $I'$ guarantee the correctness of the intent translation; the verification of $R$ and $I'$ guarantees that the issued policy complies with its underlying network constraints, and the verification of $R$ guarantees that the policy is conflict-free.
		
	\item The validity verification of $R'$ and $F$, $R$ and $F$ guarantee the practical network traffic forwarding behavior conforms to the specified policy and guarantees the validity of the intent. In the verification process, if the verification fails, the policies are corrected and reissued through policy feedback.
\end{itemize}

The intent verification framework proposed in this article realizes the full-life cycle verification of intents, which can be used as a basic framework in developing IDN.

\subsection{Use Case on Full-Life Cycle Intent-Driven Network Verification} 
	
\begin{figure}[tb]
	\centering
	\subfigure[Network topology. ]{
		\label{Fig.sub.3}
		\includegraphics[width=0.4\textwidth]{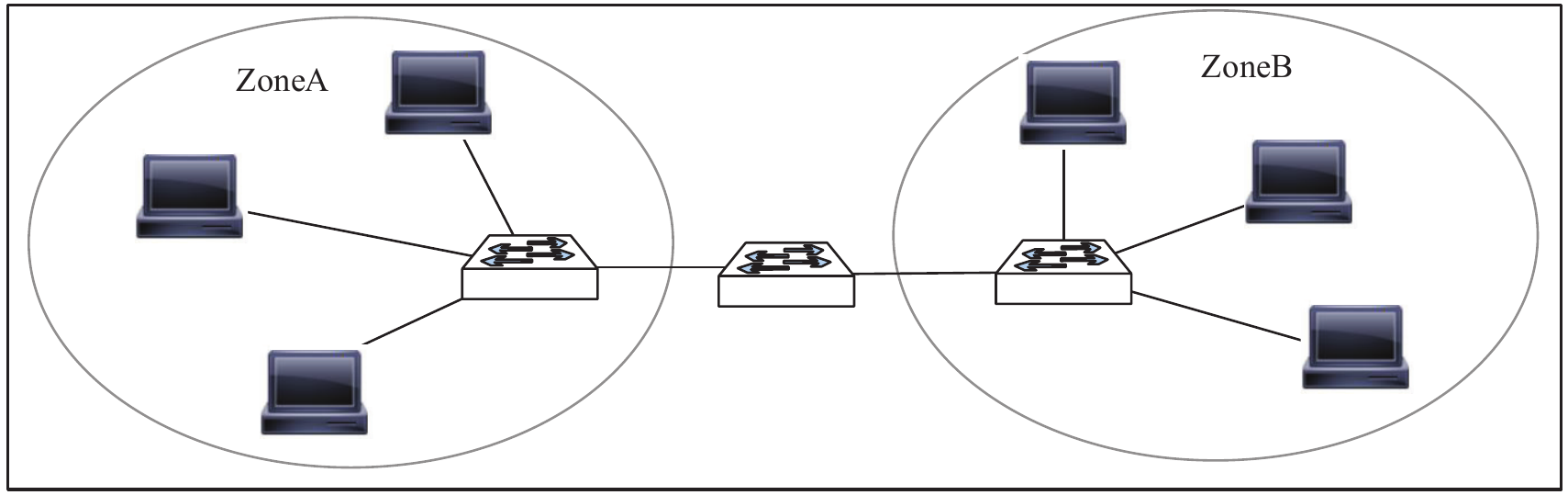}}
	\subfigure[Conflict-free policy in graph.]{
		\label{Fig.sub.1}
		\includegraphics[width=0.4\textwidth]{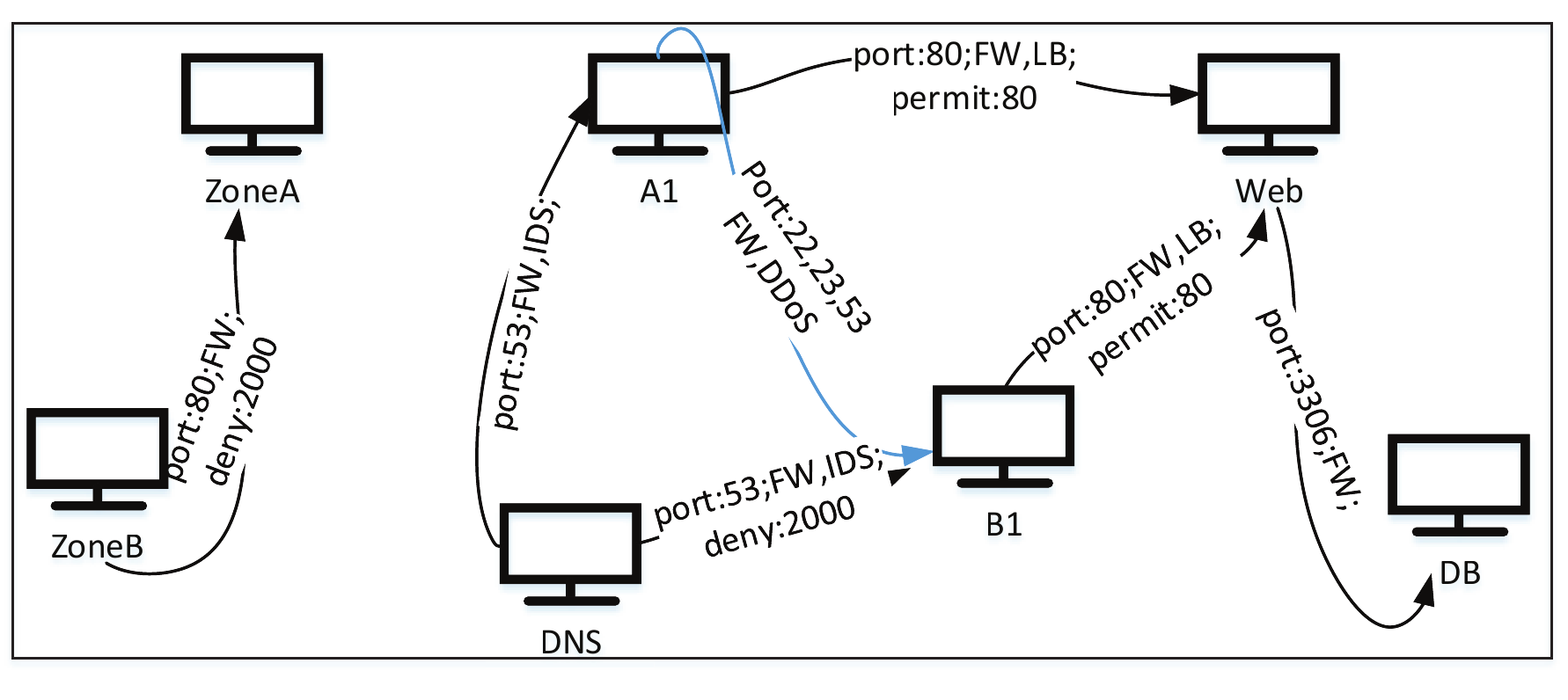}}
	\subfigure[Conflict-free policy in Pyretic. ]{
		\label{Fig.sub.2}
		\includegraphics[width=0.41\textwidth]{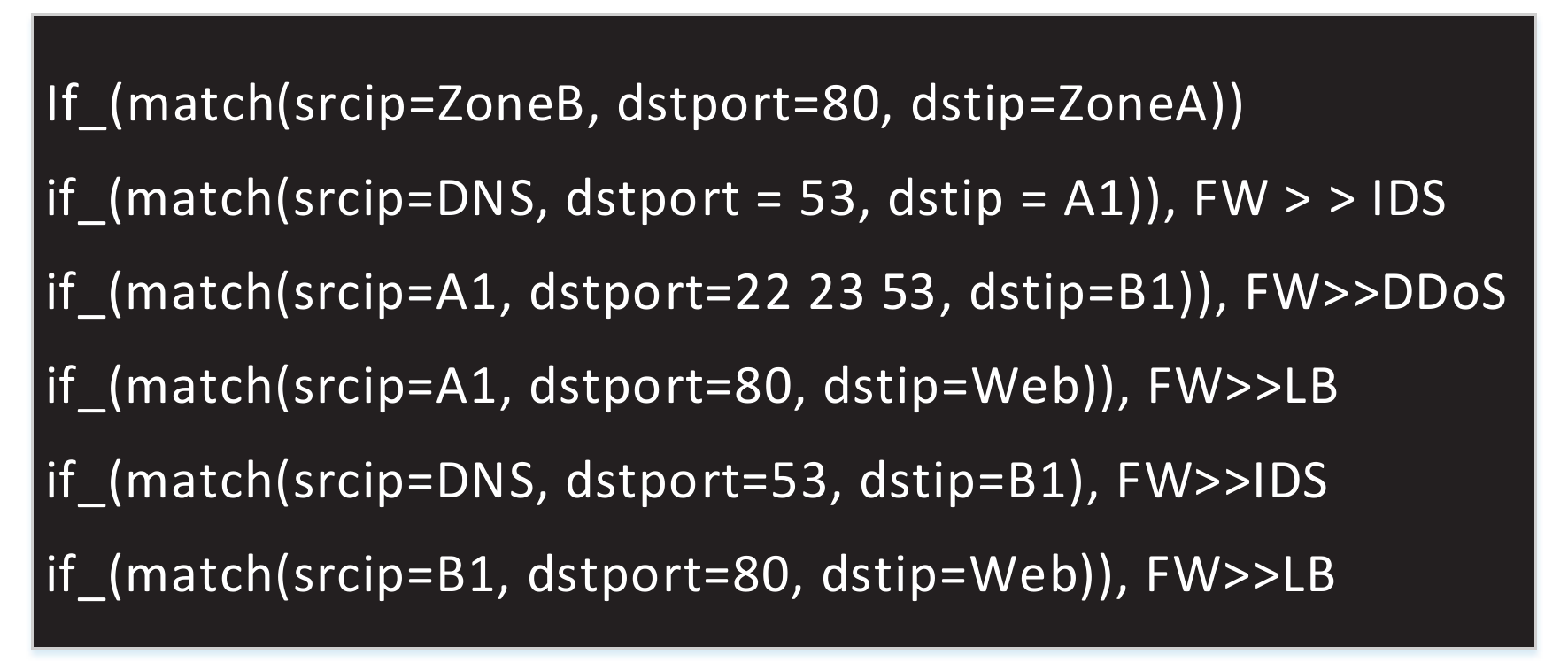}}
	\caption{The feasibility verification result: conflict-free policy in Pyretic and graph.}
	\label{Fig3.main}
\end{figure}

Network function virtualization (NFV) has become an important tool to satisfy the needs of heterogeneous services. A series of virtual network functions connected by virtual links can be used to complete the user's end-to-end service request. Therefore, we describe implementation details for realizing the proposed framework of full-life cycle intent verification outlined in the NFV simulation scenario. We represent the intent in a graph and issue the configuration into Mininet by ONOS as shown in Fig. \ref{Fig3.main} \cite{pang2020survey}. We adopt a policy graph abstraction to verify policies and convert conflict-free policies to Pyretic \cite{9580197}. Neo4j (a graph database) stores physical network information so that the intent conflict resolution module can query it. The conflict resolution algorithm is implemented in Python 2.7.0 as a single-threaded program. The network node groups in policy graph abstraction are separated based on various characteristics. All network nodes in our use case are classified based on their geographic location and function, such as Zone A and Zone B representing two buildings, A1 and B1 meaning two academies in the buildings, and world wide web (Web) and domain name system (DNS) describing different services. Finally, the policy mapper in the network controller sends the verified policy to the network. This use case can be extended to multiple network scenarios and can be integrated into other systems. It is also a beneficial research basis for implementing intent verification in future studies.

We create 100 to 500 intent application requests and the intent verification module collects the intent into $Intent \ Table \ I$. Then we realize the intent feasibility verification through graph combination. The policy graph abstraction after verification can be expressed as shown in Fig. \ref{Fig3.main} (a). Endpoints represent a group of users or network services. The text represents the port number and other attributes, such as bandwidth or network functions: load balancer ($LB$), intrusion detection systems ($IDS$), world wide web($Web$), and distributed denial of service ($DDoS$). Since the intent application requests come from different endpoint groups, they may have conflicts. At this time, the primary purpose of verification is to ensure conflict-free merging between multiple intents. The \textcolor{black}{policy graph abstraction} is translated into domain language as shown in Fig. \ref{Fig3.main} (b), which is conflict-free and executable, indicating the original address, port number, and network function between them. The service function chain planning issue is reduced to a route planning problem, with each network device representing a service function, and we select the path calculation policy from $Policy \ Table \ R$ to implement the service function.
	
\begin{figure}[ht]
			\centering
			\subfigure[The CDF of verification time for 100-500 intents.]{
				\label{fig:subfig:1} 
				\includegraphics[width=0.7\linewidth]{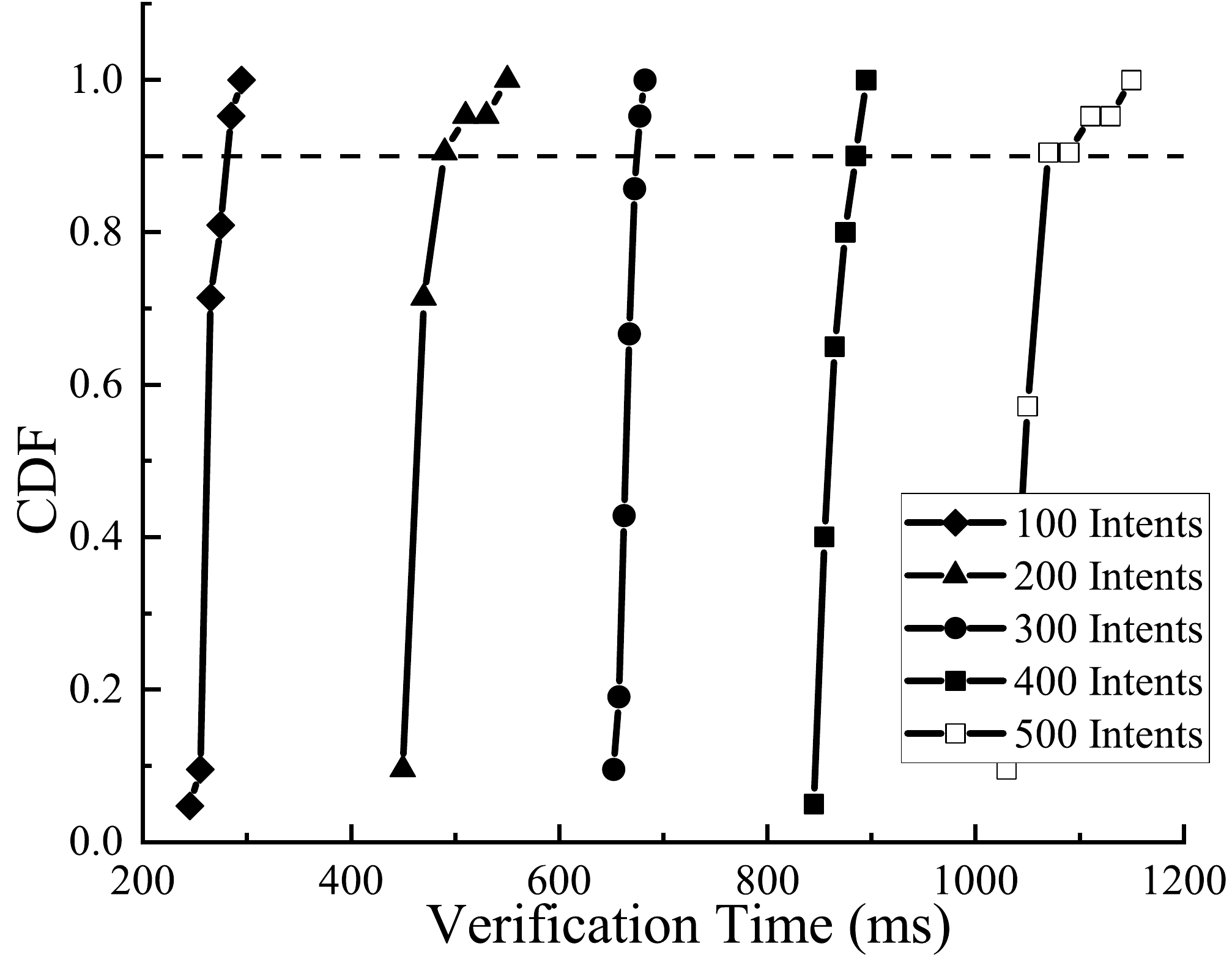}}
			\hspace{0.01in}
			\subfigure[Packet arrival rate with proposed verification engine and NIC apporach.]{
				\label{fig:subfig:6} 
				\includegraphics[width=0.7\linewidth]{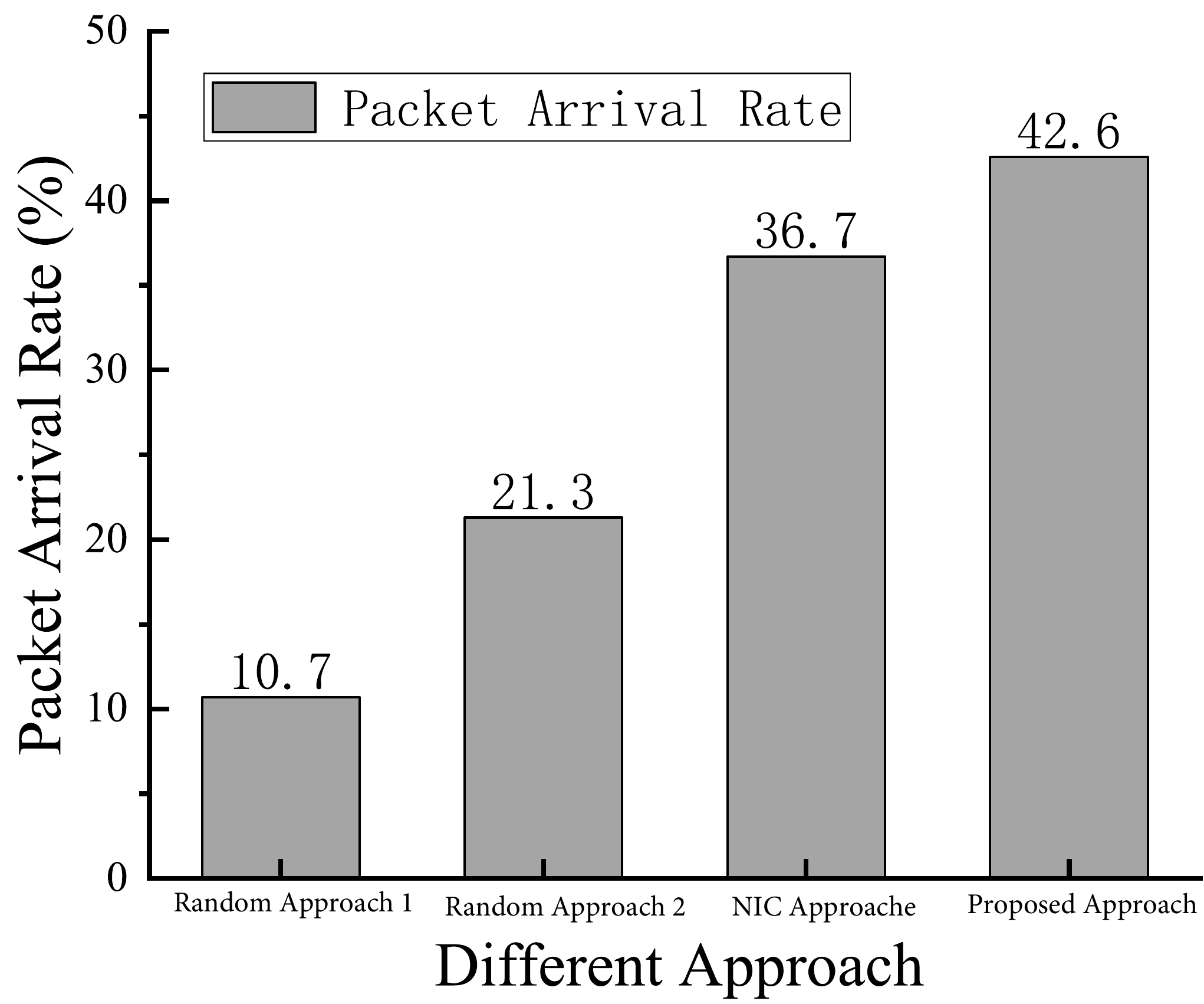}}
			\caption{The validity verification results: the verification time performance and packet arrival rate. Random approaches 1 and 2 are directly executed without processing and randomly selecting the conflicting intents to execute. NIC approach is based on network intent composition and proposed approach is our verification engine.}
			\label{fig:subfig} 
		\end{figure}

From the curves in Fig. \ref{fig:subfig} (a), when there are 100-500 intents, the verification time is between 300 and 1160 ms, and the average verification time is about 2-3 ms. The cumulative distribution function (CDF) of verification time is relatively concentrated, with a 90\% of the verification time less than 281 ms, 485 ms, 675 ms, 885 ms, and 1070 ms, respectively, which means a relatively stable effect. For the validity verification of the intent verification engine, we take the intents entered in Fig. \ref{Fig3.main} as an example, and compare the packet arrival rate with different approach. According to the simulation results in Fig. \ref{fig:subfig} (b), the worst packet arrival rate result, i.e., 10.7\% and 21.3\%, are from the approaches 1 and 2 that work by directly issuing the packets without processing, by randomly selecting one of the two conflicting intents to execution. NIC approach, i.e., based on network intent composition (NIC), yields the packet arrival rate of 36.7\%. The packet arrival rate of our verification engine is 42.6\% (approach 4). According to the results in Fig. \ref{fig:subfig}, it can be seen that our verification engine can better realize the conflict detection and decomposition of multi-user input intents.

\section{Challenge and Future Work} \label{section:5}
	
Various aspects of intent verification technology have been investigated. For example, the research has focused on formal verification, which includes formal modeling, fuzzy mathematics, and other techniques. The other research considers network device hardware and software, including controller verification servers, database development, and traffic collection techniques. However, the current technology needs to be improved in order to achieve the full-life cycle system and ensure its autonomy.
	
\begin{itemize}

\item  \textbf{The simplicity of network collection tools.} 
Continuous dynamic verification is required to combine network state. Existing network diagnostic tools are inefficient because they can assess only the forwarding behavior of a single data packet at a time. The data is obtained directly from the underlying equipment (low-level) rather than through the use of a predefined data model. Network administrators lack a global perspective on the impact of individual device modifications on the network.
		
\item  \textbf{The lack of collaboration between the network models.} Whether adopting a formal verification or a verification method based on multiple solvers, the network should be modeled, which results in limited scalability and makes it difficult to adapt to a stateful network. Validity verification includes mathematical model and traffic acquisition techniques, which require to characterize both control plane policies and data plane network states. Moreover, designing fast validation algorithms can ensure real-time validation where the mathematical model should be consistent with the feasibility validation, i.e., extend the control plane policy model to data plane modeling. There should also be modules and pipeline designs in place for the SDN controller and switches so that they can obtain and analyze data between the control plane and the data plane, as well as feedback each other.
		
\item  \textbf{The safety of intent.} 
Intents represent user preferences for networks and applications, such as demand for services, content, and network traffic. Disclosing intents may lead to leakage of user privacy. The modification of intent can affect the network more than command-line. Current verification techniques are based on the belief whether the user's intent is correct. In addition to the existing security mechanisms, the data level should be untampered with and be able to verify malicious intents. It should not only check grammar rules but semantics. For example, the intents that don't conform to network operation rules should be considered incorrect. Therefore, we should also include the security of the intent in the scope of verification. It is worth noting that current IDN also faces these challenges.
				
\end{itemize}
	
Therefore, more network verification tools and mathematical models should be the main focus in the future work, and verification techniques should be paid more attention on intent security, such as preventing intent from tampering.
	
\section{Conclusions} \label{section:6}
	
This article began by clearly defining and classifying intent-driven network verification from different perspectives, which was presented based on the location of verification, the availability of feedback, and the purpose of verification. Then we presented a brief survey of the existing verification technology. After that, we presented a full-life cycle verification framework with feedback and verified the access control policy of the network function. Our verification engine can better ensure multi-user intents conflict-free and executable. Finally, we summarized the future work and challenges of intent verification in terms of verification tools, verification models, and intent safety.

\section{Acknowledgments} \label{section:7}

This work was supported in part by the National Key Research and Development Program of China (2020YFB1807700).

\bibliographystyle{IEEEtran}
\bibliography{IEEENetwork-FullLifeCycle-YANBO-FIN}

\begin{IEEEbiographynophoto}{Yanbo Song}
	
	received his bachelor degree from Xidian University. He is currently pursuing his doctoral degree in GUIDE family at Xidian University, which is led by Dr. Chungang Yang. His research interests are intent-driven network security and network management.
	
\end{IEEEbiographynophoto}

\vspace{-12 mm}

\begin{IEEEbiographynophoto}{Chungang Yang}
	
	is a full professor at Xidian University, where he leads the research team of ~"GUIDE, Game, Utility, artificial Intelligent Design for Emerging communications~". His research interests are artificial intelligent 6G wireless mobile networks, intent-driven networks (IDN), space-terrestrial networks (STN), and game theory for emerging communication networks.
	
\end{IEEEbiographynophoto}

\vspace{-12 mm}

\begin{IEEEbiographynophoto}{Jiaming Zhang}
	
	received her master degree from Xidian University in GUIDE family, which is led by Dr. Chungang Yang. Her research interests are intent-driven network.
	
\end{IEEEbiographynophoto}

\vspace{-12 mm}

\begin{IEEEbiographynophoto}{Xinru Mi}
	received the B.E. degree in Electronic and information engineering from Northwest Normal University, Lanzhou, China. She is currently pursuing her Ph.D. degree in communication and information system. Her research interests include intent-driven network and space information network.	
\end{IEEEbiographynophoto}

\vspace{-12 mm}

\begin{IEEEbiographynophoto}{Dusit Niyato}
	is a professor in the School of Computer Science and Engineering, at Nanyang Technological University, Singapore. He received B.Eng. from King Mongkuts Institute of Technology Ladkrabang (KMITL), Thailand in 1999 and Ph.D. in Electrical and Computer Engineering from the University of Manitoba, Canada in 2008.	
\end{IEEEbiographynophoto}

\end{document}